\documentclass[letterpaper]{article}

\usepackage{aaai}
\usepackage{times}
\usepackage{helvet}
\usepackage{courier}
\usepackage{threeparttable}
\usepackage{xfrac}

\usepackage{booktabs}
\usepackage{multirow}
\usepackage{float}
\usepackage{mathtools}
\usepackage{soul}
\usepackage{color}
\usepackage{enumitem}
\usepackage{hyperref}
\usepackage{diagbox}
\usepackage{makecell}
\usepackage[toc,page]{appendix}
\usepackage[colorinlistoftodos]{todonotes}
\usepackage{subfigure}
\usepackage{algorithm}
\usepackage{algpseudocode}
\usepackage{amsmath}
\usepackage{amssymb}
\usepackage{mathtools}
\usepackage[switch]{lineno}

\begin{document}
%
\setlength\titlebox{2.6in}
\title{Deep Learning Based Detection and Localization of \\ Intracranial Aneurysms in Computed Tomography Angiography}
\author{
    Dufan Wu\textsuperscript{\rm 1, \rm 2}\thanks{Equal contribution.},
    Daniel Montes\textsuperscript{\rm 3}\footnotemark[1],
    Ziheng Duan\textsuperscript{\rm 1},
    Yangsibo Huang\textsuperscript{\rm 1},
    \AND
    Javier Romero\textsuperscript{\rm 3},
    R. Gilberto González\textsuperscript{\rm 3},
    Quanzheng Li\textsuperscript{\rm 1, \rm 2}\thanks{Corresponding author.}
    \\
    \\
    {\rm 1} Center for Advanced Medical Computing and Analysis, Massachusetts General Hospital and Harvard Medical School, Boston MA, 02114\\\\
    {\rm 2} Gordon Center for Medical Imaging, Massachusetts General Hospital and Harvard Medical School, Boston MA, 02114\\\\
    {\rm 3} Department of Radiology, Massachusetts General Hospital and Harvard Medical School, Boston MA, 02114\\
}
\vspace{0.5cm}


\maketitle
\begin{abstract}
\begin{quote}
Purpose: To develop CADIA, a supervised deep learning model based on a region proposal network coupled with a false-positive reduction module for the detection and localization of intracranial aneurysms (IA) from computed tomography angiography (CTA), and to assess our model’s performance to a similar detection network. 

Methods: In this retrospective study, we evaluated 1,216 patients from two separate institutions who underwent CTA for the presence of saccular IA $\geq$ 2.5 mm. A two-step model was implemented: a 3D region proposal network for initial aneurysm detection and 3D DenseNets for false-positive reduction and further determination of suspicious IA. Free-response receiver operative characteristics (FROC) curve and lesion-/patient-level performance at established false positive per volume (FPPV) were also performed. Fisher’s exact test was used to compare with a similar available model.

Results: CADIA’s sensitivities at 0.25 and 1 FPPV were 63.9\% and 77.5\%, respectively. Our model’s performance varied with size and location, and the best performance was achieved in IA between 5-10 mm and in those at anterior communicating artery, with sensitivities at 1 FPPV of 95.8\% and 94\%, respectively. Our model showed statistically higher patient-level accuracy, sensitivity, and specificity when compared to the available model at 0.25 FPPV and the best F-1 score (P$\le$0.001). At 1 FPPV threshold, our model showed better accuracy and specificity (P$\le$0.001) and equivalent sensitivity. 

Conclusions: CADIA outperformed a comparable network in the detection task of IA. The addition of a false-positive reduction module is a feasible step to improve the IA detection models.
\end{quote}
\end{abstract}

\section{Introduction}
Intracranial aneurysms (IA) are a complex and often multifactorial pathology where the involved mechanisms ultimately contribute to arterial wall degeneration \cite{grobelny2011brain}. 
With a prevalence of 3-6\% among the general population \cite{vlak2011prevalence,schievink1997intracranial}, IA are known for having a wide range of prognostic possibilities, ranging from asymptomatic disease to neurological disability and death \cite{ucas2012natural}. 
Aneurysmal rupture is seen in 15-30\% of IA \cite{muehlschlegel2018subarachnoid} and carries a considerably high long-term neurological disability, evidenced in up to 35\% of these patients \cite{muehlschlegel2018subarachnoid}.
By identifying high-risk IA, clinicians can reduce the risk for initial or recurrent subarachnoid hemorrhage (SAH) \cite{johnston2002recommendations}.

Although the gold standard for diagnosing IA is digital subtraction angiography (DSA) \cite{hirai2001preoperative}, it is used with caution due to its invasive nature and high risk of complications.
Indirect luminal imaging techniques such as computed tomography angiography (CTA) and magnetic resonance angiography (MRA) are more commonly used in clinical practice \cite{huston1994blinded,yang2017small}.
In the acute setting, the preferred modality is usually CTA due to easier accessibility, faster turnover times, and good accuracy in ruling out other time-sensitive pathology.
However, the interpretation of CTA is time-consuming and requires training, and a high false-negative rate in IA detection can be anticipated in clinical practice, especially among trainees \cite{hochberg2011accuracy,thompson2015guidelines}. 

Artificial intelligence (AI) permeates many aspects of modern society.
Recently, machine learning techniques have been implemented in the field of medical imaging as they excel in object recognition \cite{lecun2015deep}.
Previous algorithms in the medical imaging field revolve around breast imaging, lung nodule recognition, brain hemorrhage, among many others \cite{oren2020artificial}.
Similarly, AI-derived algorithms have been used to help increase the accuracy of aneurysm detection in clinical practice, with overall promising results \cite{park2019deep,ueda2019deep,shi2020clinically,dai2020deep}.
Therefore, the deployment of AI in aneurysm detection tasks is projected to improve the need for higher accuracy in this field. 

In this manuscript, we propose CADIA (Computer Assisted Detection of Intracranial Aneurysm), a supervised deep learning model with a false-positive reduction (FPR) module trained for the detection and localization of IA.
The model’s generalizability was investigated by testing the model on both external sites and newly acquired data.
Additionally, we compared our model’s performance to a similar detection network, DLCA \cite{yang2021deep}.
To our knowledge, this is the first time such detection + FPR algorithm has been proposed for this clinical task in the literature.

\section{Materials and Methods}
\subsection{Patient selection}
This retrospective study was approved by our Institutional Review Board and waived the requirement for patient consent.
Two retrospective searches were conducted using a centralized clinical data registry.
Our search included patients from two institutions for a non-continuous timeframe of 13 years.
Our inclusion criteria encompassed patients who underwent head or head and neck CTA and presented with one or more IA 2.5 $\geq$ mm in size, with or without SAH.
Patients with (1) one or more IA $\le$ 2.5 mm or infundibular dilations, (2) fusiform and pseudo-aneurysms, (3) vascular malformations, and (4) previous IA ligation (coil and/or clip) or embolization were excluded from our dataset. 

\subsection{Data collection}
The cohort’s demographic information was similarly retrieved using two centralized clinical data registries.
The collected demographical data included sex, race, and age. 

\subsection{Image acquisition}
CTAs were performed using multi-slice scanners by Siemens (Siemens Healthineers, Erlangen, Germany) and GE (General Electric Medical System, Chicago, USA), Phillips (Phillips Healthcare, Amsterdam, Netherlands), and Toshiba (Toshiba Medical, Tochigi, Japan).
The parameters were: 80–140 kVp, 170–350 mA, 18–25 cm field of view (FOV), and 0.6–1.25 mm slice thickness.
A power injector (rate 4-5 mL/second) canalized into an antecubital vein was used to administer iodinated contrast.
The arterial phase was achieved using SmartPrep, a semi-automatic bolus triggering technique.
Post-processing techniques were performed to include multi-planar reformats and maximum intensity projection. 

\subsection{Image evaluation}
Saccular IAs were defined as focal outpouchings $\geq$ 2.5mm in the intracranial vasculature that did not compromise the entire vessel circumference, whereas infundibular dilations were defined as triangular or funnel-shaped dilations $\le$ 2.5mm that lacked an aneurysmal neck \cite{saltzman1959infundibular}.
All CTA studies were evaluated for IA by a post-doctoral research fellow with 3 years of experience in neuroradiology.
Previous reports were available to corroborate imaging findings.
In case of discrepancy between the image evaluation and the previous report, a consensus was reached with a practicing neuroradiologist with 20 years of experience.
The evaluated vessel segments included the intracranial portion of the internal carotid artery (ICA), the anterior cerebral artery (ACA), the middle cerebral artery (MCA), the posterior cerebral artery (PCA), the basilar artery (BA), the anterior communicating artery (ACOM), and the posterior communicating artery (PCOM).
The location, size, number, and presence of SAH were recorded. 

\subsection{Model development}
We used a two-step model for IA detection: a 3D region proposal network \cite{ren2015faster} (RPN) was first trained to locate the IA, then 3D DenseNets \cite{huang2017densely} was trained to classify at each suspicious location if an IA presents.
The two-step approach drastically reduces the difficulty of training a high-performance detection network, which needs to sample from the whole volume during training and has a much longer training time. 

Before feeding the CTA volumes to the networks, each volume was truncated to at most 20cm from the cranial side to exclude regions below the neck.
The image pixel values were then cropped to [-1000, 1000] HU and scaled to [-1, 1]. 

We adopted the Dual-Pass Network (DPN) \cite{zhu2018deeplung} for the detection step.
It has a UNet structure \cite{ronneberger2015u} composed of the “Dual-Pass Blocks”, which combines dense connections and residual learning.
The DPN takes 96×96×96 patches as input due to the GPU memory limit
On each patch, it predicts the probability and bounding box of IAs inside the patch on a 24×24×24 grid.
The predicted bounding boxes with high probabilities were preserved and the overlapped ones were filtered by non-maximum suppression.

After the DPN was trained, suspicious IA locations were selected on the images by setting the model to maximum sensitivity.
At each suspicious location, three patches with different sizes (20×20×10, 32×32×16, 48×48×32) were extracted and a DenseNet was trained to predict the probability of IA for each patch \cite{dou2016multilevel}.
The probabilities predicted from the three different patch sizes were averaged as the final IA probability. 

Both steps were trained on the training dataset while monitoring the performance on the validation dataset.
The performance on the validation dataset was used to select hyperparameters and stopping criteria.
More details of the model structures and training are given in supplemental material and Figure \ref{figureE1}. 

\subsection{Model validation}
The model was validated on data acquired after the training data dates and from both internal and external sites.
Free-response receiver operative characteristics (FROC) curve (lesion-level sensitivity versus FPPV) was primarily used for the model evaluation.
An IA is counted as being found if any predicted bounding box has its center located inside the IA.
We also calculated the averaged lesion-level sensitivity at certain FPPVs. 

Besides lesion-level performance, we also evaluated the patient-level performance by predicting the existence of IA in each patient.
The models were evaluated with ROC curves, sensitivity, specificity, etc.
Fisher’s exact test was also used to test if the performance of CADIA and DLCA were statistically different for this task.

\begin{figure*}[t]
\centering    
\includegraphics[width=6.5in]{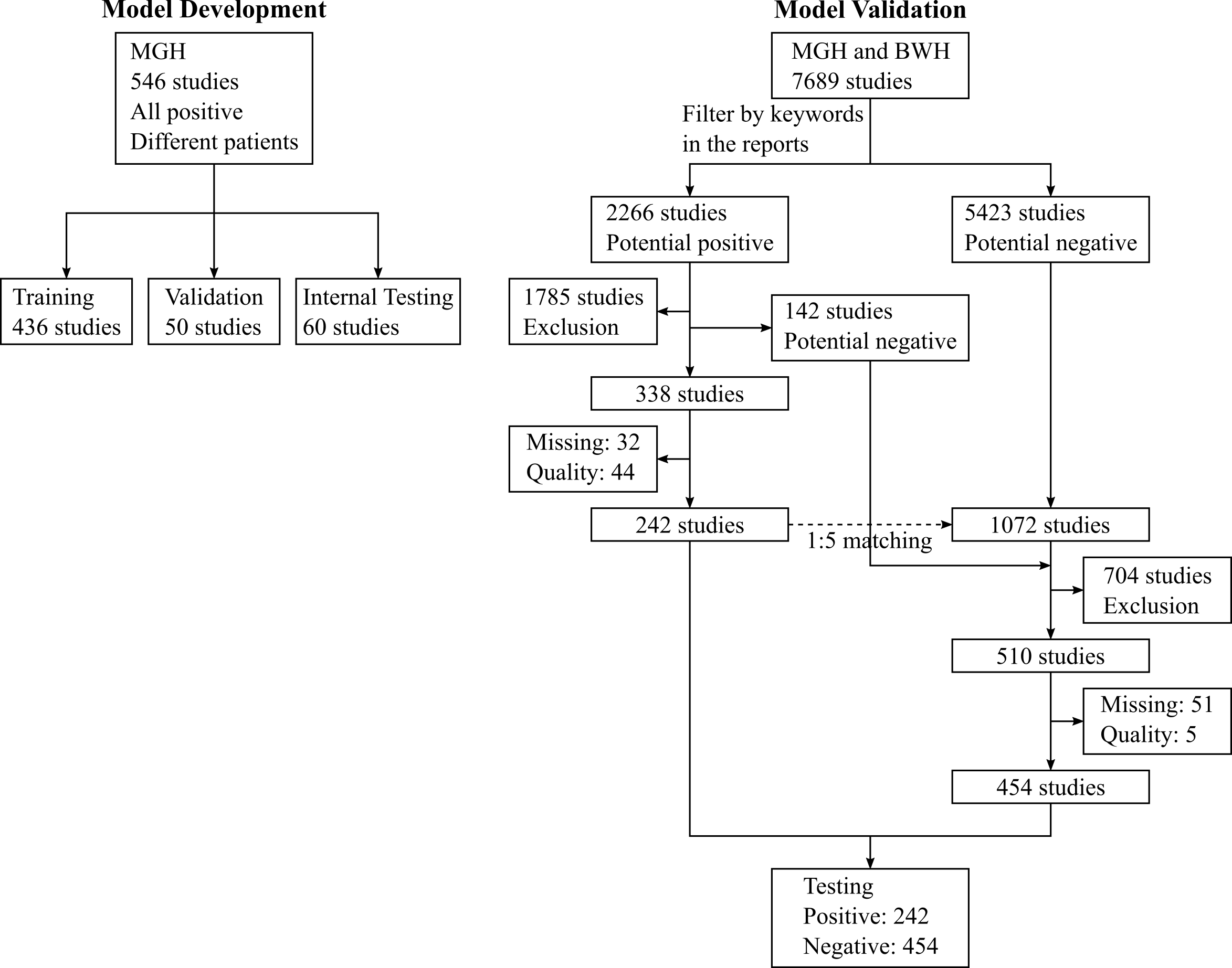} 
\caption{The data flowchart. The model was developed on a repository dataset. It was tested on data retrieved from the patient scan database at MGH and BWH. Radiology reports from these studies were first analyzed to select eligible ones, which were then downloaded from the PACS for further annotation and analysis. To reduce the huma burden to select negative cases, 1072 negative candidates were first selected by matching to the positive studies on scanning year, patient’s gender and age, where 1 positive study matches at most 5 negative studies.}
\label{figure1}
\vspace{-0.2cm}
\end{figure*}

\section{Results}
\subsection{Protocol distribution}
This is a retrospective multi-institutional study and years of inclusions spanned over two decades.
Consequently, protocols and scanners changed across the cohorts.
The data used for model development was derived from a previously published cohort \cite{heit2016detection} from Massachusetts General Hospital (MGH) and the scanner distribution was predominantly composed by (1) GE Healthcare: 422 studies (96.8\%); and (2) Siemens: 14 studies (3.2\%).
The tube potential in this cohort ranged from 120-140 KVP, the majority of them were done at 140 KVP (64.4\%).
For the testing set, the studies came from both MGH and Brigham and Women's Hospital (BWH), and the scanner distribution is as follows: (1) Siemens: 436 studies (62.6\%); (2) GE: 233 studies (33.5\%); (3) Phillips: 17 studies (2.5\%); and (4) Toshiba: 10 studies (1.4\%).
The tube potential in this cohort ranged from 80-140 KVP, and the majority of them were done at 100 KVP (81.8\%).

\subsection{Patient selection}
During our study period, we identified a total of 5,299 potential CTA exams for eligibility, composed of 3,033 exams for model development and 2,266 for model validation.
A total of 4,292 were excluded for the following reasons: presence of $\geq$ 1 infundibular dilation or IAs $\le$ 2.5mm (52\%), previously treated IA (29\%), pseudoaneurysms or fusiform aneurysms (13\%), suboptimal contrast opacification or significant venous contamination (4\%), vascular malformations including brain or dural arteriovenous fistulas (1\%), and motion degraded (1\%).
The total number of IAs across all cohorts was 872, which were divided into training, internal testing, validation, and testing set. 

In the training set, a total of 472 IAs from 436 scans from 436 patients (152 males; mean age 57.1 years) were included in our study.
In the testing set, a total of 696 scans from 670 patients were included, of which, 242 scans presented with a total of 280 IAs.
The validation set was composed of 56 IAs from 50 patients (14 males; mean age 57.7 $\pm$ 13.3 years), while the internal testing set had 64 IAs from 60 patients (19 males; mean age 57.3 $\pm$ 13.6 years).
A detailed workflow diagram for patient selection can be seen in Figure \ref{figure1}. 

\begin{table*}[]
\centering
\renewcommand\arraystretch{1.2}
\caption{Characteristics of the dataset.}
\begin{threeparttable}
\begin{tabular}{ccccc}
\hline
\multicolumn{1}{l}{\textbf{Characteristics}}     & \multicolumn{1}{l}{\textbf{Training}} & \multicolumn{1}{l}{\textbf{Validation}} & \multicolumn{1}{l}{\textbf{Internal   Testing}} & \multicolumn{1}{l}{\textbf{Testing}} \\ \hline
\multicolumn{1}{l}{\textbf{No. of patients}}     & 436                                   & 50                                      & 60                                              & 670                                  \\
\multicolumn{1}{l}{\textbf{Gender}}              &                                       &                                         &                                                 &                                      \\
Male (\%)                                        & 119 (27.3)                            & 14 (28.0)                               & 19 (31.7)                                       & 217 (32.4)                           \\
Female (\%)                                      & 317 (72.7)                            & 36 (72.0)                               & 41 (68.3)                                       & 453 (67.6)                           \\
\multicolumn{1}{l}{\textbf{Mean Age (±SD)}}      & 57.1 (±14.1)                          & 57.7 (±13.3)                            & 57.3 (±13.6)                                    & 64.9 (±14.9)                         \\
\multicolumn{1}{l}{\textbf{No. of studies}}      & 436                                   & 50                                      & 60                                              & 696                                  \\
\multicolumn{1}{l}{\textbf{No. of studies with}} &                                       &                                         &                                                 &                                      \\
0 aneurysm                                       & 0 (0.0)                               & 0 (0.0)                                 & 0 (0.0)                                         & 454 (65.2)                           \\
1 aneurysm                                       & 400 (91.7)                            & 44 (88.0)                               & 56 (93.3)                                       & 210 (30.2)                           \\
$\geq$ 2 aneurysms                                    & 36 (8.3)                              & 6 (12.0)                                & 4 (6.7)                                         & 32 (4.6)                             \\
\multicolumn{1}{l}{\textbf{SAH}}                 &                                       &                                         &                                                 &                                      \\
SAH+ (\%)                                        & 239 (54.8)                            & 30 (60.0)                               & 26 (43.3)                                       & 295 (42.4)                           \\
SAH- (\%)                                        & 197 (45.2)                            & 20 (40.0)                               & 34 (56.7)                                       & 401 (57.6)                           \\
\multicolumn{1}{l}{\textbf{Site}}                &                                       &                                         &                                                 &                                      \\
MGH (\%)                                      & 436 (100.0)                           & 50 (100.0)                              & 60 (100.0)                                      & 424 (60.9)                           \\
BWH (\%)                                      & 0 (0.0)                               & 0 (0.0)                                 & 0 (0.0)                                         & 272 (39.1)                           \\
\multicolumn{1}{l}{\textbf{Study Year}}          &                                       &                                         &                                                 &                                      \\
2002–2010 (\%)                                   & 436 (100.0)                           & 50 (100.0)                              & 60 (100.0)                                      & 0 (0.0)                              \\
2015–2020 (\%)                                   & 0 (0.0)                               & 0 (0.0)                                 & 0 (0.0)                                         & 696 (100.0)                          \\
\multicolumn{1}{l}{\textbf{KVP}}                 &                                       &                                         &                                                 &                                      \\
80 (\%)                                          & 0 (0.0)                               & 0 (0.0)                                 & 0 (0.0)                                         & 14 (2.0)                             \\
100 (\%)                                         & 0 (0.0)                               & 0 (0.0)                                 & 0 (0.0)                                         & 569 (81.8)                           \\
120 (\%)                                         & 155 (35.6)                            & 18 (36.0)                               & 20 (33.3)                                       & 112 (16.1)                           \\
140 (\%)                                         & 281 (64.4)                            & 32 (64.0)                               & 40 (66.7)                                       & 1 (0.1)                              \\
\multicolumn{1}{l}{\textbf{Manufacturer}}        &                                       &                                         &                                                 &                                      \\
GE Healthcare (\%)                               & 422 (96.8)                            & 48 (96.0)                               & 59 (98.3)                                       & 233 (33.5)                           \\
Siemens (\%)                                     & 14 (3.2)                              & 2 (4.0)                                 & 1 (1.7)                                         & 436 (62.6)                           \\
Philips (\%)                                     & 0 (0.0)                               & 0 (0.0)                                 & 0 (0.0)                                         & 17 (2.5)                             \\
Toshiba (\%)                                     & 0 (0.0)                               & 0 (0.0)                                 & 0 (0.0)                                         & 10 (1.4)                             \\
\multicolumn{1}{l}{\textbf{Total aneurysms}}     & 472                                   & 56                                      & 64                                              & 280                                  \\
\multicolumn{1}{l}{\textbf{Aneurysm size}}       &                                       &                                         &                                                 &                                      \\
2.5-3 mm (\%)                                    & 68 (14.4)                             & 5 (8.9)                                 & 7 (10.9)                                        & 101 (36.1)                           \\
3–5 mm (\%)                                      & 200 (42.4)                            & 24 (42.9)                               & 27 (42.2)                                       & 98 (35.0)                            \\
5–10 mm (\%)                                     & 168 (35.6)                            & 22 (39.3)                               & 25 (39.1)                                       & 72 (25.7)                            \\
\textgreater{}10 mm (\%)                         & 36 (7.6)                              & 5 (8.9)                                 & 5 (7.8)                                         & 9 (3.2)                              \\
\multicolumn{1}{l}{\textbf{Aneurysm location\tnote{*}}}  &                                       &                                         &                                                 &                                      \\
ICA (\%)                                         & 99 (21.0)                             & 11 (19.6)                               & 9 (14.1)                                        & 65 (23.2)                            \\
MCA (\%)                                         & 98 (20.8)                             & 17 (30.4)                               & 19 (29.7)                                       & 72 (25.7)                            \\
PCOM (\%)                                        & 77 (16.3)                             & 11 (19.6)                               & 8 (12.5)                                        & 60 (21.4)                            \\
PCA (\%)                                         & 11 (2.3)                              & 1 (1.8)                                 & 4 (6.2)                                         & 10 (3.6)                             \\
BA (\%)                                          & 35 (7.4)                              & 4 (7.1)                                 & 4 (6.2)                                         & 13 (4.6)                             \\
ACOM (\%)                                        & 122 (25.8)                            & 11 (19.6)                               & 18 (28.1)                                       & 50 (17.9)                            \\
ACA (\%)                                         & 30 (6.4)                              & 1 (1.8)                                 & 2 (3.1)                                         & 10 (3.6)                             \\ \hline
\end{tabular}
 \begin{tablenotes}
    \footnotesize
    \centering
    \item[*] ICA - internal carotid artery;
    MCA - middle cerebral artery; \\
    PCOM - posterior communicating artery;
    PCA - posterior cerebral artery; \\
    BA - basilar artery; ACOM - anterior communicating artery; ACA - anterior cerebral artery.  
  \end{tablenotes}
\end{threeparttable}
\label{table 1}
\end{table*}

\subsection{Imaging findings}
Across all cohorts, the IA size distribution is as follows: (1) 2.5-2.9mm: 181 IAs (2) 3-5mm: 349 IAs; (3) 5-10mm: 287 IAs; and (4) $\ge$ 10mm: 55 IAs.
Similarly, the vessel distribution is as follows: 184 IAs were identified in the internal carotid artery (ICA), 206 in the middle cerebral artery (MCA), 201 in the anterior communicating artery (ACOM), 156 in the posterior communicating artery (PCOM), 56 in the tip of the basilar artery (BA), 43 in the anterior cerebral artery (ACA), and 26 in the posterior cerebral artery (PCA).
SAH was present in 590 scans.
A more detailed breakdown of aneurysms composing each model can be seen in Table \ref{table 1}.

\begin{figure}[t]
\centering    
\includegraphics[scale=0.8]{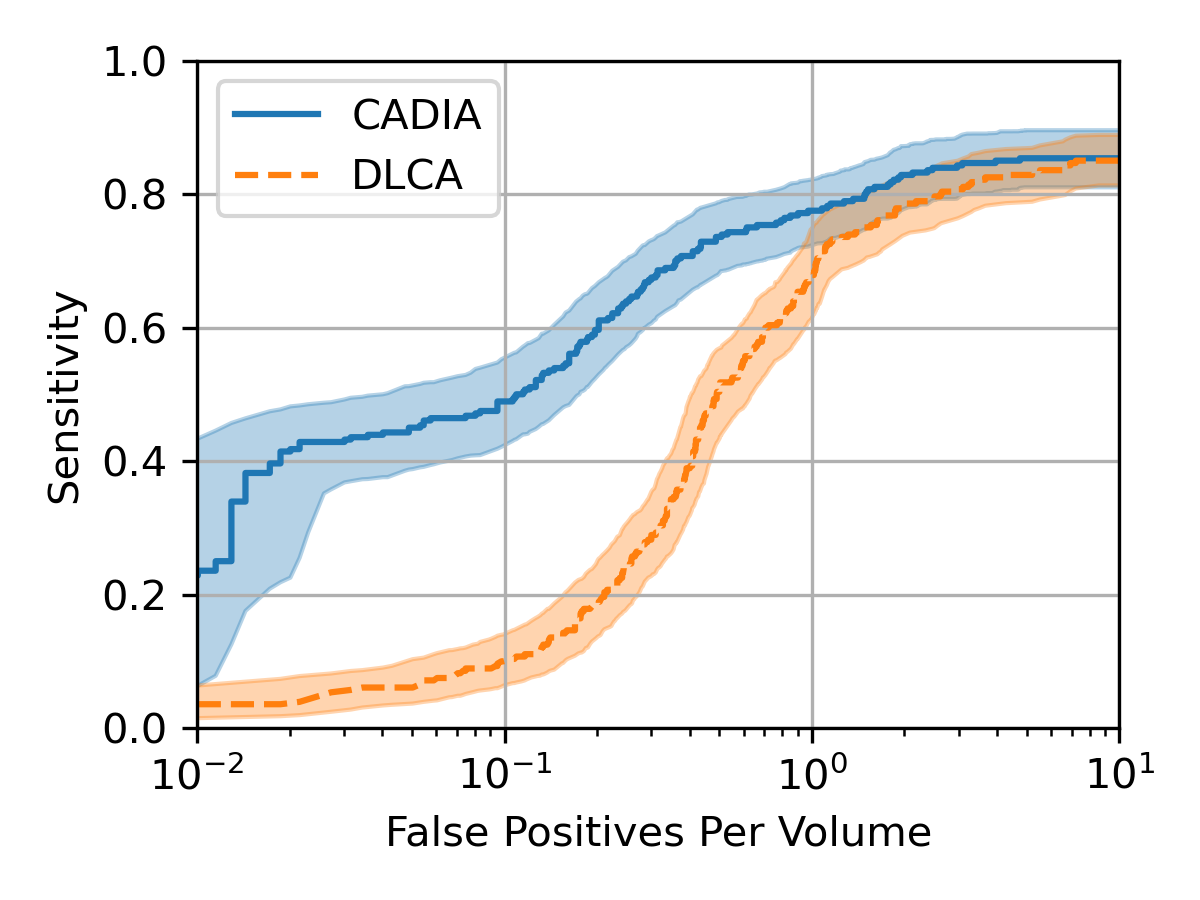} 
\caption{The FROC curve of the proposed CADIA compared to DLCA. The averaged sensitivity at FPPV = 0.125, 0.25, 0.5, 1, 2, 4, 8 for CADIA and DLCA were 0.741 (95\% CI: 0.694 – 0.784) and 0.571 (95\% CI: 0.530 – 0.608) respectively. 95\% CI is also displayed, which was calculated using 1,000 times of bootstrapping.}
\label{figure2}
\vspace{-0.2cm}
\end{figure}

\begin{figure}[t]
\centering    
\includegraphics[scale=0.8]{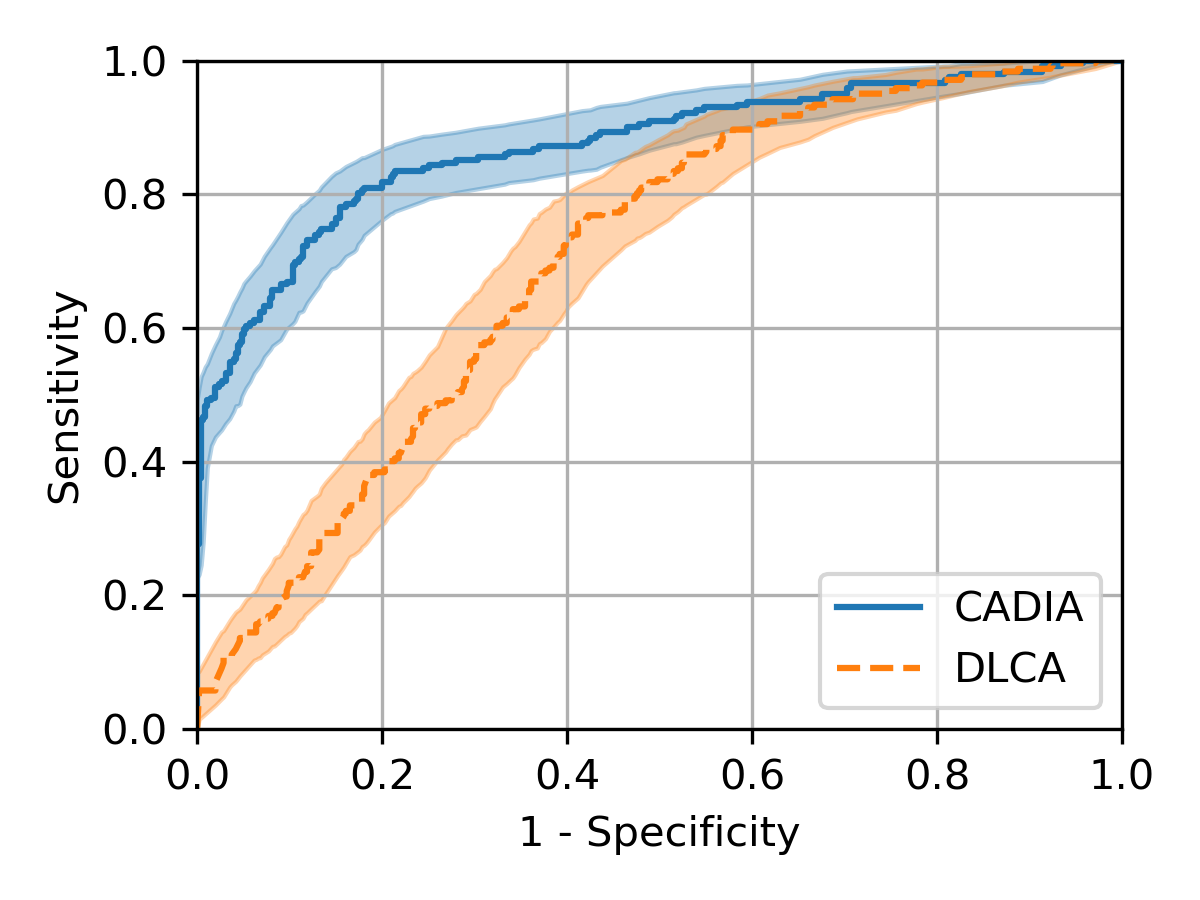} 
\caption{The volume-level ROC curves of the proposed CADIA and DLCA. A volume is (predicted) positive if it contains one or more (predicted) aneurysm(s). The AUCs for CADIA and DLCA were 0.873 (95\% CI: 0.843 – 0.901) and 0.702 (95\% CI: 0.663 – 0.740). 95\% CI is also displayed, which was calculated using 1,000 times of bootstrapping.}
\label{figure3}
\vspace{-0.2cm}
\end{figure}

\begin{figure}[t]
\centering    
\includegraphics[scale=0.55]{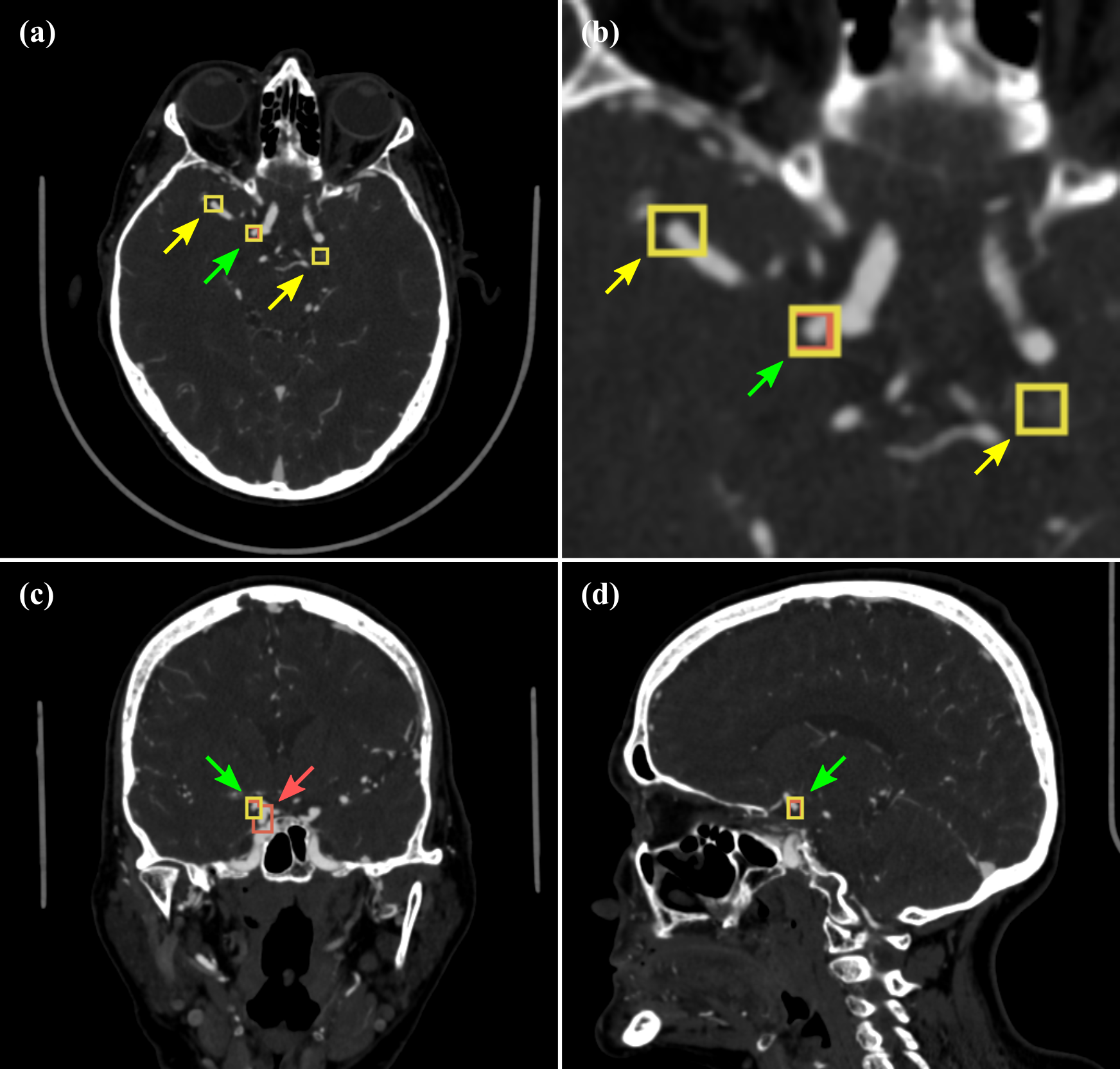} 
\caption{CADIA’s performance at 1 FPPV on a 65-year-old female patient with multiple aneurysms. The model’s predicted aneurysms are marked by yellow boxes whereas the annotations are marked by red boxes. The model captured a 3mm aneurysm located in the MCA (green arrow), which was not captured by the DLCA. Both models missed the 6mm aneurysm (red arrow) adjacent to the 3mm one. CADIA had two false negatives (yellow arrows) due to local vessel shape (left) and veins.}
\label{figure4}
\vspace{-0.2cm}
\end{figure}

\begin{figure}[t]
\centering    
\includegraphics[scale=0.55]{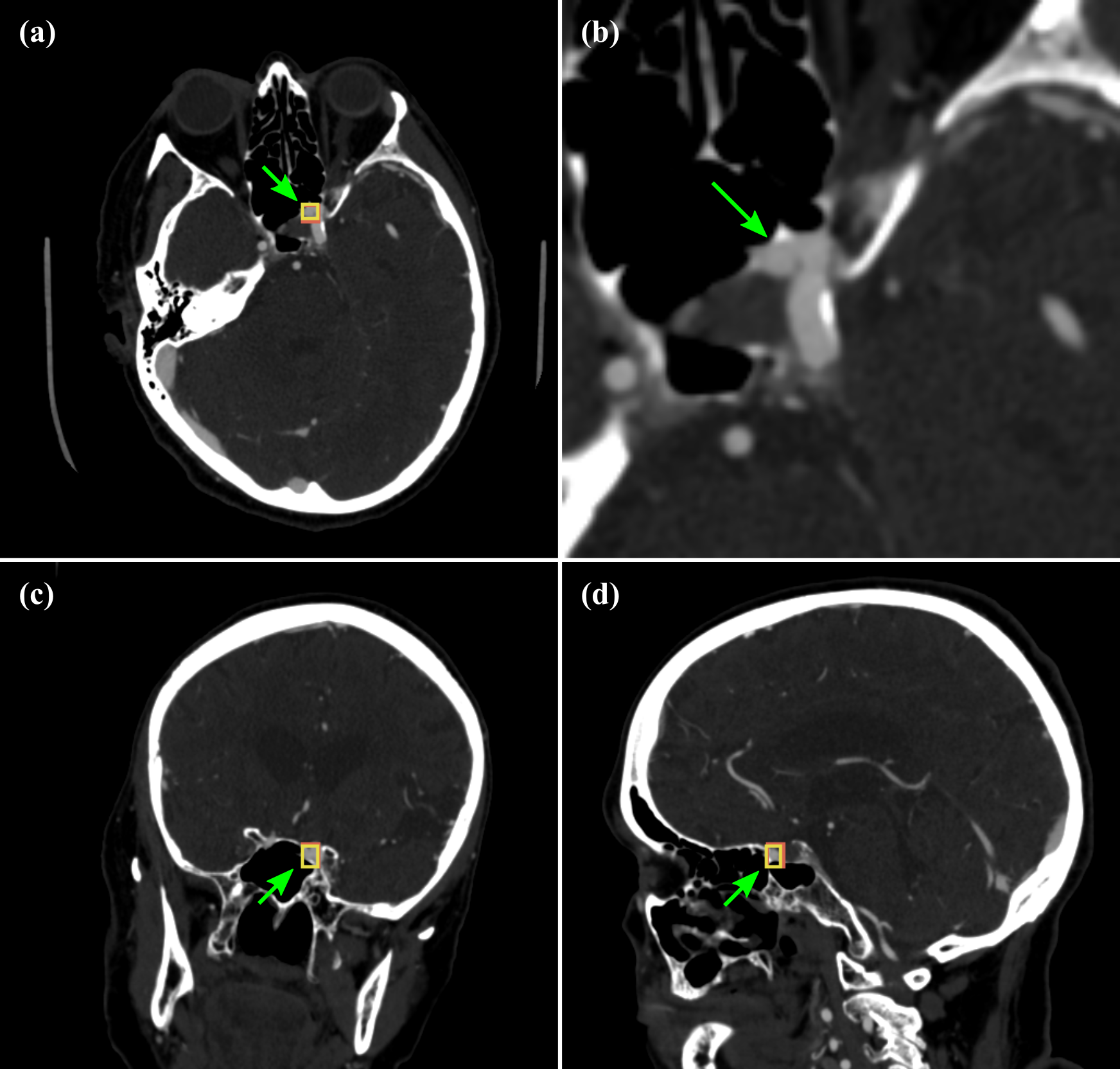} 
\caption{The volume-level ROC curves of the proposed CADIA and DLCA. A volume is (predicted) positive if it contains one or more (predicted) aneurysm(s). The AUCs for CADIA and DLCA were 0.873 (95\% CI: 0.843 – 0.901) and 0.702 (95\% CI: 0.663 – 0.740). 95\% CI is also displayed, which was calculated using 1,000 times of bootstrapping.}
\label{figure5}
\vspace{-0.2cm}
\end{figure}

\begin{table*}[]
\renewcommand\arraystretch{1.2}
\caption{The stratified performance of the two models on the testing set at 0.25 and 1 FPPV.}
\centering
\begin{tabular}{lllll}
\hline
\multirow{2}{*}{\textbf{Target}}   & \multicolumn{2}{c}{\textbf{Sensitivity   at 0.25 FPPV {[}95\% CI{]}}}                           & \multicolumn{2}{c}{\textbf{Sensitivity   at 1 FPPV {[}95\% CI{]}}}                              \\ \cline{2-5} 
                                   & \textbf{CADIA}                                 & \textbf{DLCA}                                  & \textbf{CADIA}                                 & \textbf{DLCA}                                  \\ \hline
\textbf{All}                       & \multicolumn{1}{c}{0.639 {[}0.582,   0.694{]}} & \multicolumn{1}{c}{0.243 {[}0.193,   0.293{]}} & \multicolumn{1}{c}{0.775 {[}0.725,   0.824{]}} & \multicolumn{1}{c}{0.679 {[}0.625,   0.730{]}} \\
\textbf{Aneurysm size}             & \multicolumn{1}{c}{}                           & \multicolumn{1}{c}{}                           & \multicolumn{1}{c}{}                           & \multicolumn{1}{c}{}                           \\
\multicolumn{1}{c}{$\leq$3 mm (n=101)}  & \multicolumn{1}{c}{0.406 {[}0.314,   0.505{]}} & \multicolumn{1}{c}{0.090 {[}0.036,   0.153{]}} & \multicolumn{1}{c}{0.594 {[}0.495,   0.690{]}} & \multicolumn{1}{c}{0.520 {[}0.430,   0.618{]}} \\
\multicolumn{1}{c}{3-5 mm (n=98)}  & \multicolumn{1}{c}{0.663 {[}0.564,   0.755{]}} & \multicolumn{1}{c}{0.212 {[}0.135,   0.292{]}} & \multicolumn{1}{c}{0.816 {[}0.736,   0.895{]}} & \multicolumn{1}{c}{0.677 {[}0.585,   0.769{]}} \\
\multicolumn{1}{c}{5-10 mm (n=72)} & \multicolumn{1}{c}{0.917 {[}0.852,   0.973{]}} & \multicolumn{1}{c}{0.444 {[}0.326,   0.556{]}} & \multicolumn{1}{c}{0.958 {[}0.905,   1.000{]}} & \multicolumn{1}{c}{0.875 {[}0.795,   0.947{]}} \\
\multicolumn{1}{c}{$\geq$10 mm (n=9)}   & \multicolumn{1}{c}{0.778 {[}0.429,   1.000{]}} & \multicolumn{1}{c}{0.667 {[}0.333,   1.000{]}} & \multicolumn{1}{c}{0.889 {[}0.667,   1.000{]}} & \multicolumn{1}{c}{0.889 {[}0.625,   1.000{]}} \\
\textbf{Aneurysm location}         & \multicolumn{1}{c}{}                           & \multicolumn{1}{c}{}                           & \multicolumn{1}{c}{}                           & \multicolumn{1}{c}{}                           \\
\multicolumn{1}{c}{ICA (n=65)}     & \multicolumn{1}{c}{0.415 {[}0.304,   0.533{]}} & \multicolumn{1}{c}{0.092 {[}0.028,   0.169{]}} & \multicolumn{1}{c}{0.600 {[}0.482,   0.716{]}} & \multicolumn{1}{c}{0.600 {[}0.486,   0.707{]}} \\
\multicolumn{1}{c}{MCA (n=72)}     & \multicolumn{1}{c}{0.764 {[}0.662,   0.857{]}} & \multicolumn{1}{c}{0.347 {[}0.231,   0.468{]}} & \multicolumn{1}{c}{0.889 {[}0.808,   0.957{]}} & \multicolumn{1}{c}{0.667 {[}0.559,   0.764{]}} \\
\multicolumn{1}{c}{PCOM (n=60)}    & \multicolumn{1}{c}{0.583 {[}0.459,   0.714{]}} & \multicolumn{1}{c}{0.233 {[}0.125,   0.351{]}} & \multicolumn{1}{c}{0.717 {[}0.589,   0.828{]}} & \multicolumn{1}{c}{0.717 {[}0.597,   0.820{]}} \\
\multicolumn{1}{c}{PCA (n=10)}     & \multicolumn{1}{c}{0.600 {[}0.250,   0.909{]}} & \multicolumn{1}{c}{0.100 {[}0.000,   0.333{]}} & \multicolumn{1}{c}{0.600 {[}0.250,   0.909{]}} & \multicolumn{1}{c}{0.500 {[}0.166,   0.833{]}} \\
\multicolumn{1}{c}{BA (n=13)}      & \multicolumn{1}{c}{0.462 {[}0.182,   0.765{]}} & \multicolumn{1}{c}{0.231 {[}0.000,   0.500{]}} & \multicolumn{1}{c}{0.769 {[}0.500,   1.000{]}} & \multicolumn{1}{c}{0.846 {[}0.631,   1.000{]}} \\
\multicolumn{1}{c}{ACOM (n=50)}    & \multicolumn{1}{c}{0.860 {[}0.756,   0.957{]}} & \multicolumn{1}{c}{0.320 {[}0.196,   0.448{]}} & \multicolumn{1}{c}{0.940 {[}0.863,   1.000{]}} & \multicolumn{1}{c}{0.800 {[}0.682,   0.902{]}} \\
\multicolumn{1}{c}{ACA (n=10)}     & \multicolumn{1}{c}{0.700 {[}0.375,   1.000{]}} & \multicolumn{1}{c}{0.300 {[}0.000,   0.616{]}} & \multicolumn{1}{c}{0.800 {[}0.500,   1.000{]}} & \multicolumn{1}{c}{0.400 {[}0.091,   0.714{]}} \\
\textbf{Site}                      & \multicolumn{1}{c}{}                           & \multicolumn{1}{c}{}                           & \multicolumn{1}{c}{}                           & \multicolumn{1}{c}{}                           \\
\multicolumn{1}{c}{MGH (n=424)} & \multicolumn{1}{c}{0.604 {[}0.530,   0.678{]}} & \multicolumn{1}{c}{0.213 {[}0.157,   0.273{]}} & \multicolumn{1}{c}{0.740 {[}0.673,   0.805{]}} & \multicolumn{1}{c}{0.621 {[}0.551,   0.695{]}} \\
\multicolumn{1}{c}{BWH (n=272)} & \multicolumn{1}{c}{0.694 {[}0.596,   0.784{]}} & \multicolumn{1}{c}{0.288 {[}0.206,   0.373{]}} & \multicolumn{1}{c}{0.829 {[}0.748,   0.903{]}} & \multicolumn{1}{c}{0.766 {[}0.685,   0.839{]}} \\
\textbf{SAH}                       & \multicolumn{1}{c}{}                           & \multicolumn{1}{c}{}                           & \multicolumn{1}{c}{}                           & \multicolumn{1}{c}{}                           \\
\multicolumn{1}{c}{SAH+ (n=295)}   & \multicolumn{1}{c}{0.667 {[}0.581,   0.748{]}} & \multicolumn{1}{c}{0.252 {[}0.175,   0.328{]}} & \multicolumn{1}{c}{0.784 {[}0.704,   0.856{]}} & \multicolumn{1}{c}{0.649 {[}0.564,   0.735{]}} \\
\multicolumn{1}{c}{SAH- (n=401)}   & \multicolumn{1}{c}{0.621 {[}0.550,   0.695{]}} & \multicolumn{1}{c}{0.237 {[}0.175,   0.303{]}} & \multicolumn{1}{c}{0.769 {[}0.698,   0.839{]}} & \multicolumn{1}{c}{0.698 {[}0.633,   0.765{]}} \\ \hline
\end{tabular}
\label{table 2}
\end{table*}

\begin{table*}[]
\centering
\renewcommand\arraystretch{1.2}
\caption{The patient-level performance of the two models at different thresholds.}
\begin{threeparttable}
\begin{tabular}{ccccc}
\hline
\multicolumn{1}{l}{\textbf{}}               & \multicolumn{1}{l}{\textbf{Accuracy}} & \multicolumn{1}{l}{\textbf{Sensitivity}} & \multicolumn{1}{l}{\textbf{Specificity}} & \multicolumn{1}{l}{\textbf{F1-Score}} \\ \hline
\multicolumn{1}{l}{\textbf{0.25 FPPV}}      & \multicolumn{1}{l}{}                  & \multicolumn{1}{l}{}                     & \multicolumn{1}{l}{}                     & \multicolumn{1}{l}{}                  \\
CADIA                                       & 0.828 {[}0.797,   0.855{]}            & 0.723 {[}0.664,   0.776{]}               & 0.883 {[}0.854,   0.914{]}               & 0.745 {[}0.699,   0.786{]}            \\
DLCA                                        & 0.657 {[}0.624,   0.690{]}            & 0.430 {[}0.370,   0.494{]}               & 0.778 {[}0.742,   0.820{]}               & 0.465 {[}0.409,   0.518{]}            \\
P-Value                                     & \textless 0.001                       & \textless 0.001                          & \textless 0.001                          &                                       \\
\multicolumn{1}{l}{\textbf{1 FPPV}}         &                                       &                                          &                                          &                                       \\
CADIA                                       & 0.678 {[}0.644,   0.716{]}            & 0.893 {[}0.855,   0.932{]}               & 0.564 {[}0.519,   0.612{]}               & 0.659 {[}0.616,   0.700{]}            \\
DLCA                                        & 0.547 {[}0.510,   0.586{]}            & 0.921 {[}0.883,   0.954{]}               & 0.348 {[}0.303,   0.392{]}               & 0.586 {[}0.544,   0.628{]}            \\
P-Value                                     & \textless 0.001                       & 0.347                                    & \textless 0.001                          &                                       \\
\multicolumn{1}{l}{\textbf{Best F1-Score\tnote{*}}} &                                       &                                          &                                          &                                       \\
CADIA                                       & 0.816 {[}0.787,   0.843{]}            & 0.806 {[}0.757,   0.851{]}               & 0.822 {[}0.788,   0.857{]}               & 0.753 {[}0.709,   0.793{]}            \\
DLCA                                        & 0.605 {[}0.570,   0.639{]}            & 0.855 {[}0.810,   0.898{]}               & 0.471 {[}0.425,   0.517{]}               & 0.601 {[}0.557,   0.642{]}            \\
P-Value                                     & \textless 0.001                       & 0.182                                    & \textless 0.001                          &                                       \\ \hline
\end{tabular}
 \begin{tablenotes}
    \footnotesize
    \centering
    \item[*] The thresholds were selected using the best F1-score for each model. The FP/Image for CADIA and DLCA were 0.379 (95\% CI: 0.312 – 0.443) and 0.701 (95\% CI: 0.635 – 0.766), respectively. The p-values were calculated using Fisher’s exact test.  
  \end{tablenotes}
\end{threeparttable}
\label{table 3}
\end{table*}

\subsection{Model’s performance and comparison}
The lesion-level performance of CADIA and DLCA are shown in the FROC curves in Figure \ref{figure2}.
Their sensitivities were also measured at 0.25 (high specificity) and 1 (high sensitivity) FPPV on various subsets and given in Table 2. 

To evaluate the models’ overall performance, the averaged sensitivities at 0.125, 0.25, 0.5, 1, 2, 4, and 8 FPPV were calculated, which were 74.1\% and 57.1\% for CADIA and DLCA, respectively.
At 0.25 FPPV, the sensitivities of the CADIA and DLCA were 63.9\% and 24.3\%, respectively; at 1 FPPV, the sensitivities were 77.5\% and 67.9\%. CADIA demonstrated improved sensitivity compared to DLCA especially at lower FPPV.  

As expected, both models’ performance changed with IA size and location.
At 1 FPPV, CADIA performed the best on IAs between 5 – 10 mm and those located in ACOM, with sensitivities of 95.8\% and 94.0\%, respectively.
Aneurysms between 2.5 – 3 mm and those located in the ICA had the least favorable performance, with sensitivities of 59.4\% and 60.0\%, respectively.
The presence of SAH and sites did not significantly alter out model’s performance.
A similar conclusion can be reached for 0.25 FPPV.

Compared to DLCA, CADIA had the greatest improvement on hard cases at 0.25 FPPV, i.e., IAs $\leq$ 3mm and those located in ICA. 
Despite that the two models had similar performance for these two IA subsets at 1 FPPV, CADIA kept a sensitivity of 40.6\% for $\leq$ 3mm IA at 0.25 FPPV, whereas DLCA only had 9.0\% sensitivity.
Similarly, for IAs at ICA, the two models had close performance at 1 FPPV, but CADIA and DLCA’s sensitivities were 41.5\% and 9.2\% at 0.25 FPPV. 

The models were further evaluated for volume-level binary classification of the existence of IAs.
The ROC curves are given in Figure \ref{figure3}, and the accuracy, sensitivity, specificity, and F1-score at fixed thresholds are given in Table \ref{table 3}.

Thanks to the improved sensitivity at lower FPPV, CADIA had improved volume-level sensitivity at high specificity compared to DLCA, as shown in Figure \ref{figure3}.
The areas under the curve (AUC) of CADIA and DLCA were 0.873 and 0.702, respectively. 

The fixed thresholds analysis results were given at 0.25 FPPV, 1 FPPV and the best F1-score for each model.
CADIA had reasonable specificities and sensitivities at 0.25 FPPV (88.3\% and 72.3\%) and best F1-score (82.2\% and 80.6\%), but low specificity despite of the high sensitivity at 1 FPPV (56.4\% and 89.3\%).
Compared to CADIA, DLCA had relatively low specificity except for 0.25 FPPV (77.8\%), where its sensitivity was too low (43.0\%).
Under Fisher’s exact test, CADIA had statistically better accuracy, sensitivity, and specificity compared to DLCA at all three thresholds (P $\le$ 0.001) except for the sensitivities at 1 FPPV (P = 0.347) and best F1-score (P = 0.182).
Figure 4 and 5 show typical CTA volumes analyzed by CADIA with examples of true positives, false positives and false negatives. 

\section{Discussion}
In this study, we developed CADIA, a supervised deep learning algorithm with a false-positive reduction module for the detection and localization of IA.
Our results showed a worthy diagnostic performance that changed with aneurysm size and location, as expected \cite{yang2017small,pradilla2013accuracy}.
Additionally, we compared the performance of our model to a similar detection model, DLCA \cite{dai2020deep}, outperforming it in the vast majority of tested categories on our dataset. 

Compared to segmentation-based models such as HeadXNet14 and DLIA \cite{shi2020clinically}, CADIA is a detection network like DLCA.
It requires only bounding box labels for training instead of semantic segmentation labels, which significantly reduces the labor needed for annotation.
Both models’ maximum sensitivities were lower compared to the reported value by Yang et al \cite{dai2020deep}, which was mainly due to the difficulty of the testing dataset.
Figure \ref{figureE2} shows the FROC curves of CADIA and DLCA on the internal testing dataset (see Table \ref{table 1}), which were 92.1\% and 93.7\% respectively, which were comparable to that reported by Yang et al.
Furthermore, an ablation study on the FPR module was done and the results are shown in Figure E3, which demonstrated that the FPR module could not only improve the performance of the DPN, but also the DLCA model which was trained on a completely different dataset. 

We found that IAs between 2.5 and 3 mm, and those located in the cavernous segment of the ICA and PCA are the most difficult for our model to identify.
We attribute the subpar performance in these locations to the low prevalence of PCA aneurysms in our training set (2\%), and to the vessel tortuosity and close relationship to bony structures in the ICA \cite{philipp2017comparison}.
Interestingly, our model had an outstanding performance in the ACOM complex, one of the locations considered problematic for radiologists in one of these studies \cite{pradilla2013accuracy}.
Remarkably, CADIA showed improvement in sensitivity for IA detection in those studies with SAH at low FPPV thresholds, reaching a comparable sensitivity to that of trained neuroradiologists \cite{philipp2017comparison}.

The risk of IA rupture is multifactorial and involves both demographic and imaging characteristics \cite{yuan2020hemodynamic,brinjikji2017intracranial,wermer2007risk}.
A risk-additive feature whose evaluation is often heterogeneous is aneurysmal growth. A previous systematic review conducted by Brinjikji et al. \cite{brinjikji2017intracranial} evidenced that growth rates $\geq$ 2 millimeters in an IA’s maximal dimension can increase the risk of rupture as much as 3.1\% per year, compared to 0.1\% rupture risk in stable IAs. We believe that the implementation of detection algorithms in clinical settings can lessen the inter and intra-reader variability by offering a standardized approach for aneurysm follow-up. 

As previous research in this specific field has demonstrated \cite{park2019deep,ueda2019deep,shi2020clinically,dai2020deep}, we believe that the deployment of such algorithms will help physicians identify IAs and their implementation can have a constructive impact in various segments of the healthcare industry.
First, it is well known that the correctly identification of intracranial aneurysms in clinical setting is directly related to the reader’s experience \cite{hochberg2011accuracy,thompson2015guidelines}.
Therefore, implementing our model in academic centers involved in resident training and in community-based hospitals where general radiologists operate can be of great benefit.
Second, our model’s usefulness could be optimized in fast-paced and high-volume environments, such as in a private practice setting.
Third, our model could also function as a warning tool, improving radiology workflow by prioritization of positive scans through a warning system.
Lastly, implementing CADIA as a supplementary layer in the interpretation of intracranial vasculature can improve the peer-review process in radiology \cite{donnelly2007performance}.
This could offer greater significance in developing countries, where sequential interpretation of medical imaging is less usual.

The present study has some limitations.
First, the retrospective and non-consecutive nature of our study may introduce subject selection bias.
Second, the training data set did not contain an optimal number of aneurysms in certain locations, namely PCA, BA, and ACA.
Third, our model cannot provide information on rupture risk.
However, we believe that the strengths of this study overcome the deficiencies.
We consider that our study has a pronounced external validity, as we included studies from different institutions, CT scanners, and protocols, while maintaining a remarkable diagnostic performance.
Noticeably, our testing dataset had a large shift on CTA protocols compared to the training dataset due to the big difference in the acquisition years of the training and testing datasets.
The stable performance of the model argues in favor of its generalizability against the rapidly evolving CT technology.
Additionally, we performed a direct comparison with an available network, proving large superiority on our dataset.
Furthermore, we included patients with and without SAH. 

In conclusion, CADIA demonstrated a worthy diagnostic performance in the detection and localization of intracranial aneurysms, and it demonstrated superior performance when compared to a similar model.
We proved that the addition of a False-Positive Reduction module is a feasible alternative to improving the intracranial aneurysms detection models.

\bibliographystyle{aaai}
\bibliography{ref}

{

\appendixpage
\section{Detection Network}
We adapted the DPN \cite{zhu2018deeplung} as the detection network to locate aneurysms from the 3D volume. DPN is fully convolutional network with U-Net-like structure \cite{ronneberger2015u}. The network takes 96×96×96 patches as the input, and outputs 24×24×24 grids. On each voxel it gives the probability, center, and diameter of the aneurysms in the patch. To normalize the output of the network for easier training, an anchor size l is defined as the reference size of the aneurysm. Hence, the output of the network at each grid point is:

\begin{equation}
\mathbf{t}_d=\Big(p_d,\frac{x_d-x_a}{l_k},\frac{y_d-y_a}{l_k},\frac{z_d-z_a}{l_k},log(\frac{s_d}{l_k})\Big),
\end{equation}

where $p_d$ is the probability of the aneurysm; $\left(x_d,y_d,z_d\right)$ are the centers of the aneurysm; $\left(x_a,y_a,z_a\right)$ are the coordinate of the output grid point; $s_d$ is the diameter of the aneurysm in pixels; and $l_k$ is the anchor size. 

To address for the different sizes of the aneurysms, three anchors were defined at each grid point. An anchor is a bounding box centered at the grid point with anchor size $l_k$. We used $l_k=5,\ 10,\ 20$ for the three anchors according to experience. Hence, at each output grid point, 3 vectors $\mathbf{t}_d$ will be predicted by the network with different anchor sizes $l_k$. 

The network was trained by averaging the following cross-entropy + regression loss on each anchor: 
\begin{equation}
L=-p_d^*log(p_d)-(1-p_d^*)log(1-p_d)+\lambda_{reg}p_d^*||\mathbf{t}_d-\mathbf{t}_d^*||_1,
\end{equation}
where $p_d^*$ is the labeled aneurysm probability, defined as $p_d^*=1$ if there is an aneurysm and $p_d^*=0$ if not; $\lambda_{reg}=0.5$ is the weight of the regression loss; $\mathbf{t}_d$ is defined in (1) and $\mathbf{t}_d^*$ is its correspondence for the label:
\begin{equation}
\mathbf{t}_d^*=\Big(p_d^*,\frac{x_d^*-x_a}{l_k},\frac{y_d^*-y_a}{l_k},\frac{z_d^*-z_a}{l_k},log(\frac{s_d^*}{l_k})\Big),
\end{equation}
where $\left(x_d^\ast,y_d^\ast,z_d^\ast\right)$ and $s_d^\ast$ are the center and diameter of the labeled aneurysm. Note that $\mathbf{t}_d^\ast$ are only meaningful for the anchors which have an associated labeled aneurysm, otherwise the regression loss will have zero weights because $p_d^\ast=0$.

To determine if an anchor can “see” an aneurysm ($p_d^\ast=1$), we calculated the intersection over union (IoU) of the aneurysm’s bounding box with the anchor. For each anchor, its largest IoU with all the aneurysms is calculated. If the IoU is larger than 0.5, $p_d^\ast$ will be set to 1 and $\mathbf{t}_d^\ast$ will be taken from the aneurysm with the largest IoU. If the IoU is smaller than 0.02, $p_d^\ast$ will be set to 0. Otherwise, this anchor will be excluded from the training because its overlap with the aneurysms is on the borderline. 

All the patches were cropped to [-1000, 1000] HU and further normalized to [-1, 1]. During training, patches were sampled at the aneurysms’ location and randomly from the volumes in a balanced manner. Random argumentations including shift, zoom, contrast adjustment, and noise were applied to the sampled patches. Because of the small size of aneurysm compared to the whole CT volume, there were much more negative samples compared to the positive ones, and a network with small training loss may still generate tons of false positives. Hard negative mining was applied to mitigate this problem, where in each training iteration, only the top 2 negative anchors with the largest loss was used as negative samples in each patch. 

During testing, non-maximum suppression (NMS) was used to reduce the number of bounding boxes produced by the algorithm. We used an IoU threshold of 0.25 for NMS and kept all the bounding boxes with probability $p_d$ larger than 0.25. To avoid aneurysms being on the boundary of the patches during testing, we split the volume into 96×96×96 patches with 16 pixels overlap between adjacent patches. Aneurysm candidates on individual patches were predicted first and aggregated together with NMS. 

We used stochastic gradient descent (SGD) for the training. A batch size of 16 was used. The learning rate started from 0.01 and gradually reduced to ${10}^{-4}$ at the 1,000th epoch, with a momentum of 0.9. The performance of the network was monitored on the validation dataset every 100 epochs, and we stopped the training at the 400th epoch, where the maximum sensitivity first exceeded 0.9 at 10 FPPV.

\section{False-Positive Reduction Network}
The detection network has a huge hyperparameter space, a long processing chain, and a long training time which make it very inefficient to search for optimal hyperparameters of the network. Instead of finetuning the detection network, we proposed to tune it to a point with high sensitivity and reasonable false positivity and use a secondary patch-based classification network for the false positive reduction (FPR).

The FPR network is shown in Figure \ref{figureE1}, which is the 3D version of DenseNet121 \cite{huang2017densely} with slight modifications on the growth rate and the first convolutional layer. It extracts fixe-sized patches centered at the locations given by the detection network and gives the probability of aneurysm in that patch $p_f$. To accommodate for different aneurysm sizes, we extracted three patch sizes at each location (20×20×10, 32×32×16, 48×48×32) \cite{dou2016multilevel}, which were determined by the aneurysm size distribution and experience. Three networks were trained for each patch size respectively, and the final probability of aneurysm at a location was given by the averaged probability from the three networks. 

To train the FPR networks, patches were extracted at the locations given by the detection network on the training dataset. Each patch was labeled as positive if its center resides inside aneurysm or negative otherwise. Patches were excluded from the training if the distance between the patch center and aneurysm center were less than half the patch size. These patches were excluded during training because they were too close to the aneurysm and may affect the training. 

All the patches were cropped to [-1000, 1000] HU and further normalized to [-1, 1]. Random argumentation including shift, zoom, flip, and noise were applied during the training. The networks were trained using cross-entropy loss with a batch size of 32, where all the patches were sampled with equal weights despite that there were more negative samples. Adam algorithm were used with initial learning rate of ${10}^{-4}$ and gradually reduced to ${10}^{-5}$ at the 500th epoch. The network’s performance on the validation dataset was evaluated every epoch and the checkpoint with the best specificity at 0.95 sensitivity were used for the final model. 

The patches can be extracted before training to greatly improve the IO efficiency. The detection network training took almost one week for 400 epochs on 4 RTX 2080Ti, whereas the FPR network training took 2 to 4 hours (depending on the patch size) for 500 epochs on a single GPU.

\begin{figure*}[t]
\centering    
\includegraphics[width=6.5in]{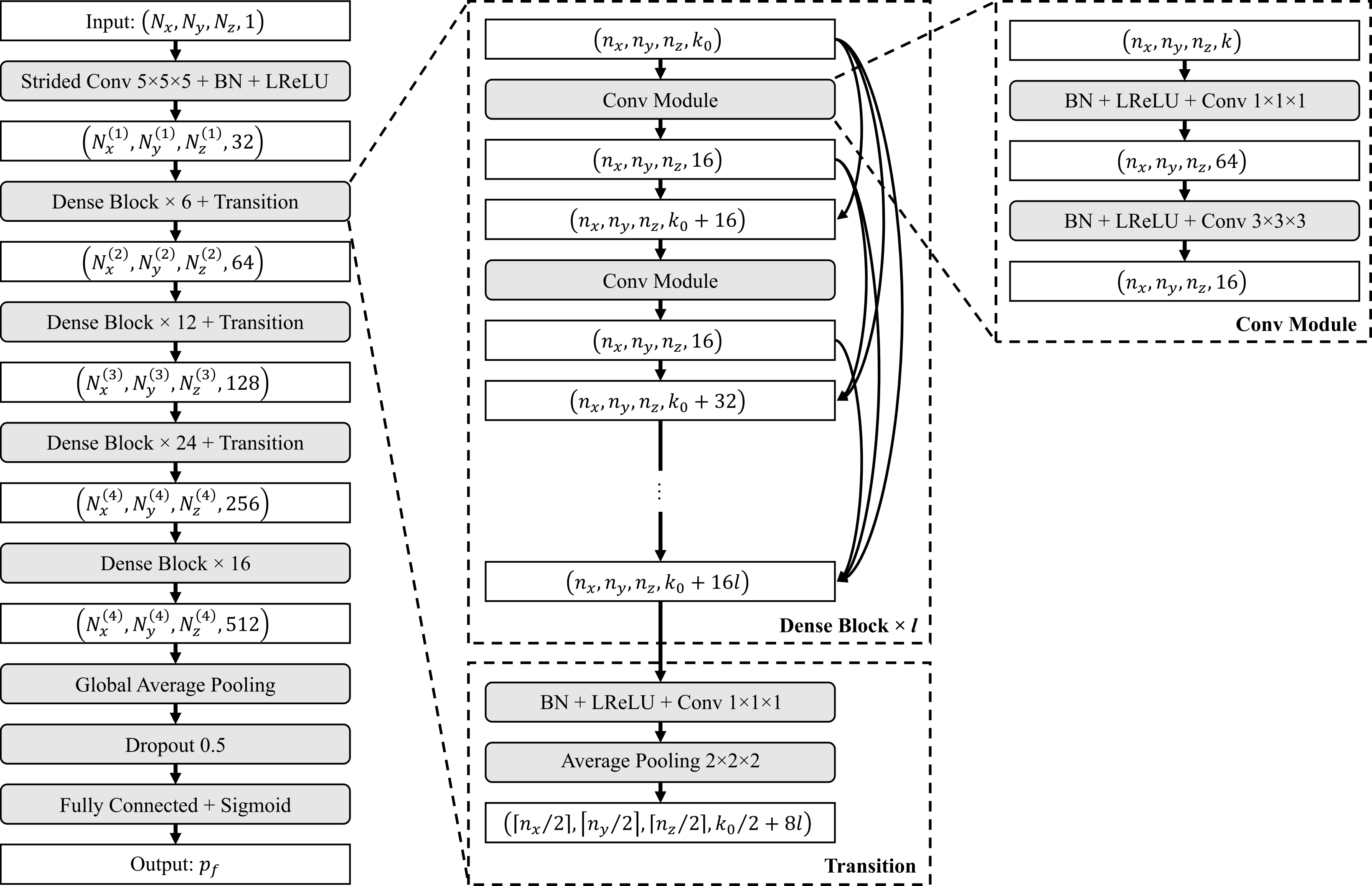} 
\caption{The structure of the 3D DenseNet121 used for FPR.
It takes an input of size $\left(N_x,N_y,N_z,1\right)$ and outputs a single probability $p_f$. The input first goes through a 5×5×5 convolutional layer with stride of 2, a batch normalization (BN) layer and a Leaky ReLU (LReLU) layer ($\alpha$=0.2).
Then it goes through 4 Dense blocks with 6, 12, 24, and 16 convolutional modules inside, where the parameters were taken from DenseNet121.
At last, a global average pooling layer is applied to make a 1-D tensor.
The tensor then goes through a dropout layer and fully connected layer to get the final predicted probability.
We used a growth rate of 16 for the dense blocks, so that given an input with $k_0$ channels to an l-layer dense block, the output will have $k_0$+16$l$ channels.
The transition module is composed of a BN + LReLU + Conv module with a 2×2×2 average pooling layer.
The convolutional module will half the channel number whereas the pooling layer will half the patch size.
We used zero padding for all the convolution and pooling operations.
Hence, the tensor size will have $N_{x,y,z}^{\left(i\right)}=\left\lceil\sfrac{N_{x,y,z}^{\left(i-1\right)}}{2}\right\rceil$, where $N_{x,y,z}^{\left(0\right)}=N_{x,y,z}$.}
\label{figureE1}
\end{figure*}

\begin{figure*}[t]
\centering    
\includegraphics[scale=1]{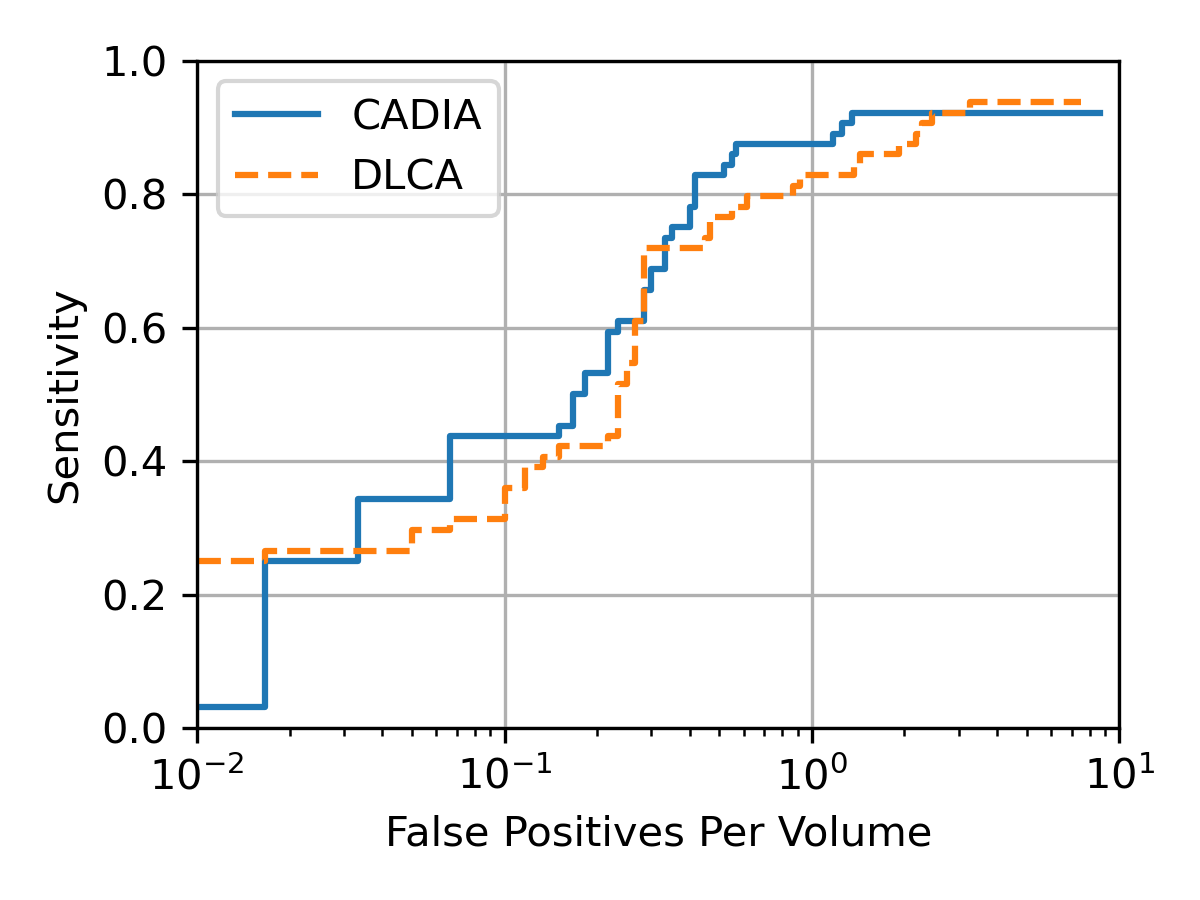} 
\caption{The FROC curves of CADIA and DLCA on the internal testing dataset, with similar distribution with the training dataset and larger aneurysm sizes compared to the testing dataset. Both models could achieve over 90\% sensitivity at maximum, which is similar to the performance reported in Yang et al.18 The maximum sensitivity of CADIA is 92.1\% (95\% CI: 85.7 – 98.3) at 8.65 FPPV, whereas that of DLCA is 93.7\% (95\% CI: 88.0 – 98.4) at 7.48 FPPV. The averaged sensitivity at FPPV = 0.125, 0.25, 0.5, 1, 2, 4, 8 for the two models are 78.7\% (95\% CI: 70.5 – 89.0) and 75.0\% (95\% CI: 66.6 – 85.0). The confidence interval is too large due to small sample size and not shown in the figure.}
\label{figureE2}
\end{figure*}

\begin{figure*}[t]
\centering    
\includegraphics[scale=1]{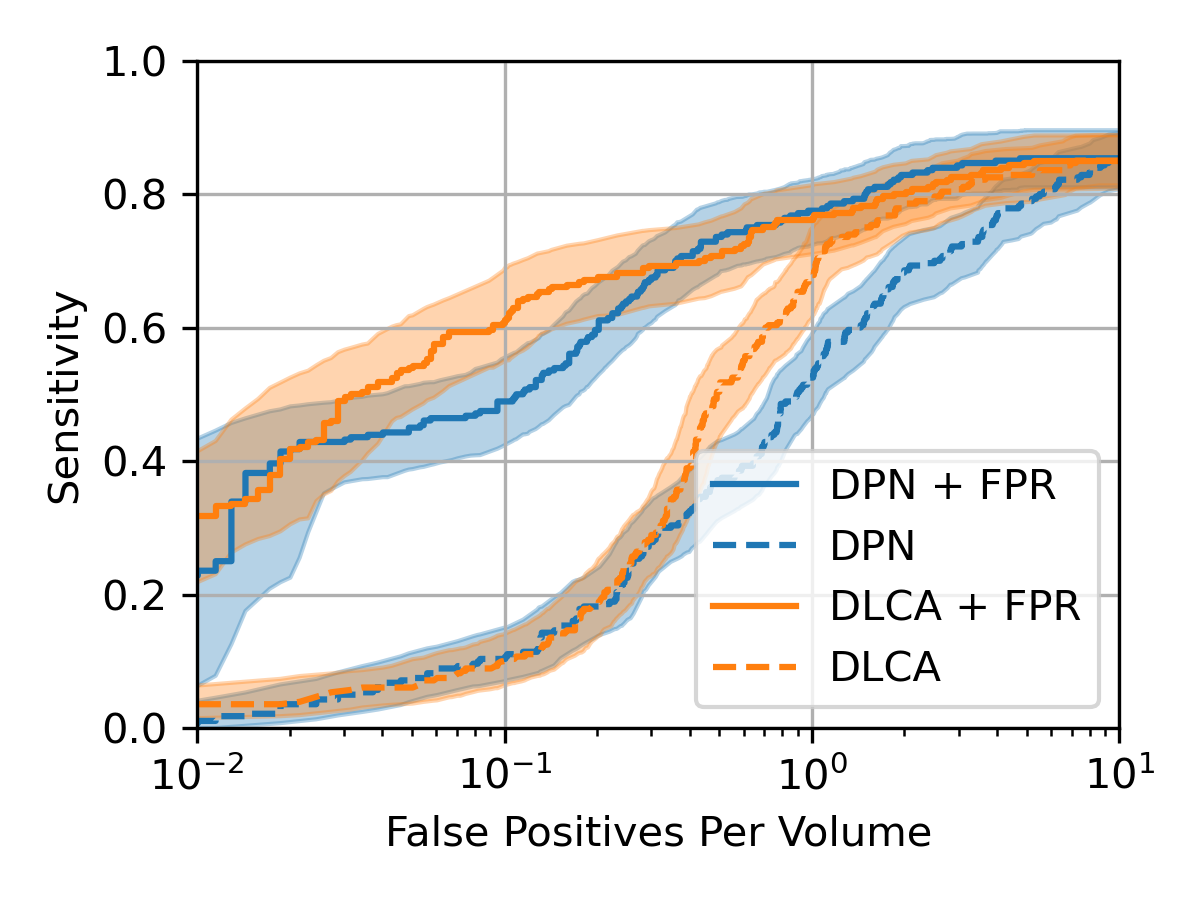} 
\caption{The FROC curves of DPN and DLCA with and without the FPR network on the testing dataset. The averaged sensitivity at FP = 0.125, 0.25, 0.5, 1, 2, 4, 8 for DPN + FPR (CADIA), DPN, DLCA + FPR, and DLCA are 0.741 (95\% CI: 0.694 – 0.784), 0.503 (95\% CI: 0.466 – 0.543), 0.754 (95\% CI: 0.710 – 0.801), and 0.571 (95\% CI: 0.530 – 0.608), respectively. When incorporating the FPR network to DLCA, we applied the trained DLCA model to the training dataset and extracted patches to train the FPR network. Isotropic patch sizes of 20×20×20, 32×32×32, and 48×48×48 were used because DLCA resampled the volume to 0.39×0.39×0.39 $mm^3$. The results showed that the FPR module could incorporate with other detection networks and further improve their performance. }
\label{figureE3}
\end{figure*}

}

\end{document}